\documentclass[fleqn,twoside]{article}
\usepackage{espcrc2}

\usepackage{epsfig}

\def\Journal#1#2#3#4{{#1} {\bf #2} (#4) #3}
\def\NPB{{\em Nucl. Phys.} B}
\def\PLB{{\em Phys. Lett.}  B}
\def\PRL{\em Phys. Rev. Lett.}
\def\PRD{{\em Phys. Rev.} D}

\def\APPB{{\em Acta. Phys. Polon.} B}

\def\IJMPA{{\em Int. J. Mod. Phys.} A}

\def\be{\begin{equation}}
\def\ee{\end{equation}}

\newcommand{\AmS}{{\protect\the\textfont2
  A\kern-.1667em\lower.5ex\hbox{M}\kern-.125emS}}

\title{Towards the lattice study of M-theory (II)\thanks{This research is supported by the
Polish Committee for Scientific Research under the grant 2 P03B
019 17.}
\thanks{Presented by J. Wosiek.}
}

\author{P. Bialas\address{Institute of Computer Science,
Jagellonian University, \\
       Nawojki 11, 30-072 Cracow, Poland}%
        and
        J. Wosiek\address{M. Smoluchowski Institute of Physics,
 Jagellonian University,\\ Reymonta 4,
30-059 Cracow, Poland }  }

\begin{document}

\begin{abstract}
We present new results of the quenched simulations of the reduced
D=4 supersymmetric Yang - Mills quantum mechanics for larger gauge
groups SU(N), $ 2<N<9 $. The model, studied at finite temperature,
reveals existence of the two distinct regions which may be
precursors of a black hole and the elementary D0 branes phases of
M-theory conjectured in the literature. Present results for higher
groups confirm the picture found already for N=2. Similar
behaviour is observed in the preliminary simulations for the D=6
and D=10 models. \vspace{1pc}
\end{abstract}

\maketitle

\section{SUPERSYMMETRIC YANG-MILLS QUANTUM MECHANICS}
     Supersymmetric Yang-Mills quantum mechanics (SYMQM) provides the
quantitative model of M-theory \cite{BFSS}. Even though much
simpler than the original theory the model is not solved in spite
of its long history\cite{CH,UPP,HS}. We have therefore decided to
set up a systematic lattice survey of SYMQM beginning with the
simplest case of $D=4, N=2, N_f=0$(quenched)\cite{JW} and
gradually extending it as far as possible towards the BFSS limit
{\em i.e.} $D=10, N\rightarrow\infty $ and $N_f=1$. In this talk I
will report on the second step along this programme: the first
results for higher N will be presented.

       The action of the SYMQM reads
\begin{equation}
 S=\int dt \left({1\over 2} \mbox{\rm Tr} F_{\mu\nu}(t)^2
                     +\bar\Psi^a(t){\cal D}\Psi^a(t) \right). \label{QM}
\end{equation}
where $\mu,\nu=1\dots D$, and all fields are independent of the
space coordinates $\vec{x}$.
 The  supersymmetric fermionic partners
 belong to the addjoint representation of SU(N). The discretized system is put
on a $D$ dimensional hypercubic lattice $N_1\times\dots\times N_D$
which is reduced in all space directions to $N_i=1$, $i=1\dots
D-1$.
 The gauge part of the action has now the usual form
\be S_G=-\beta \sum_{m=1}^{N_t} \sum_{\mu>\nu} {1\over N} Re(
\mbox{\rm Tr} \, U_{\mu\nu}(m) ), \label{SG} \ee with \be
\beta=2N/a^3 g^2,  \label{beta} \ee and $U_{\mu\nu}(m)=
U_{\nu}^{\dagger}(m)U_{\mu}^{\dagger}(m+\nu)
     U_{\nu}(m+\mu)U_{\mu}(m) $,
$U_{\mu}(m)=\exp{(iagA_{\mu}(a m))}$, where $a$ denotes the
lattice constant and $g$ is the gauge coupling in one dimension.
The integer time coordinate along the lattice is $m$. Periodic
boundary conditions $U_{\mu}(m+\nu)=U_{\mu}(m)$, $\nu=1\ldots
D-1$, guarantee that Wilson plaquettes $U_{\mu\nu}$ tend, in the
classical continuum limit, to the appropriate components
$F_{\mu\nu}$ without the space derivatives.
\section{RESULTS}
Up to date we have addressed  the two problems: 1) extracting the
continuum limit from the lattice data, and 2) the search for the
nontrivial phase structure. The first point is essential in any
approach based on the discretization. In particular restoration of
the continuum supersymmetry, of the full unquenched model, may
 crucially depend on the ability to control the continuum limit.
 The second issue is connected to the problem of the
 Bekenstein-Hawking entropy which has an elegant solution in the
 framework of M-theory \cite{STRO}. Namely the supersymmetric, extremal black holes
found in the latter can be viewed as composed of the elementary D0
brane excitations, providing the statistical interpretation of the
area of the "Schwarzschild" horizon which is known to behave as an
etropy. In particular, the theory also predicts existence of the
two phases in which the gravity and the elementary D0 branes
provide good description respectively\cite{MAR}.
\subsection{SU(2)}
We have found in \cite{JW} that the continuum limit of the model
can be readily extracted with the bare parameters scaling with
canonical dimensions. This is expected for the one dimensional
system. To search for the phase transition we have studied the
distribution of the eigenvalues of the Polyakov line
  $L=  \prod_{m=1}^{N_t} U_D(m)$,
which is a very sensitive determinant of the phase structure in
gauge theories.
 It was found that, similarly to the large volume QCD, in the low temperature phase
the eigenvalues are concentrated around zero, while at high
temperature the distribution is peaked around $\pm 1$ which
constitute the center of SU(2). In the space extended theories the
$Z_2$ symmetry is spontaneously broken in the infinite volume
limit and only one direction is populated. In the present 0-volume
system, this may happen only in the infinite-N limit, the
Gross-Witten 
 model being a known example of the
critical behaviour emerging at large N.
\begin{figure}[h]
 \epsfxsize=17pc \epsfysize=27pc
 \epsfbox{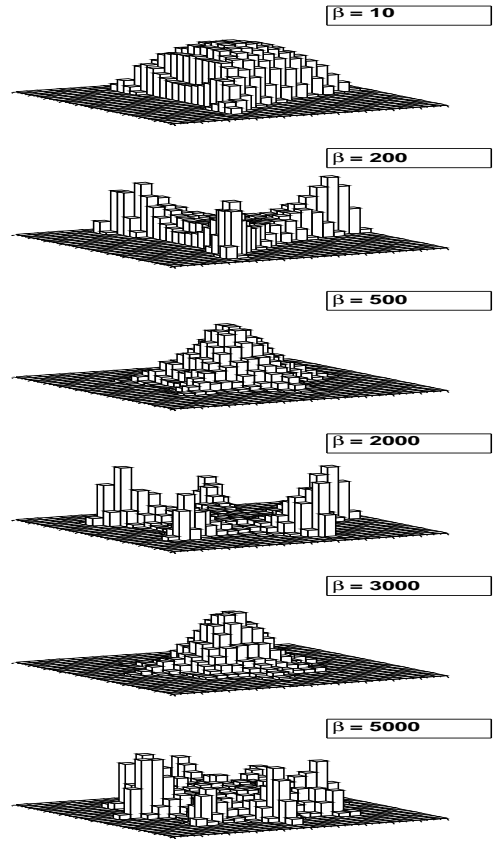}
 \caption{Distributions of the eigenvalues of $L$, at low and high
 temperatures, for N=3,5,8.} \label{fig:toosmall}
\end{figure}
\subsection{SU(3 - 8)}
For higher groups we find now the same behaviour, see Fig.1. Since
$\beta=2N N_t^3 T^3 $, the histograms correspond to the low and high
temperature regions for a range of $N$. Evidently they change from
convex to concave at some critical value of $\beta
(\equiv\beta_c)$ similary to the N=2 case. The nature of the transition
is not resolved yet.
Nevertheless, our data show unambiguously that the system behaves
differently in both regions. For example, we have also measured
the dependence of the size of the system, $R^2=g^2\sum_a
(A_i^a)^2$, on the temperature, and found that it is definitely
different, and the change in the behaviour occurs at the same T
where the distributions in Fig. 1 change their shapes. It is also
possible, for the first time to confront the N dependence of the
transition temperature with theoretical expectations. It follows
from Eq.(\ref{beta}) and $N_t=1/Ta$, that the 't Hooft scaling
$T_c\approx (g^2 N)^{1/3}$ implies that the lattice coupling
$\beta_c\approx N^2$ at fixed $N_t$. Fig.2 shows the $N$
dependence of $\beta_c/N^2$ for available range of N together with
the fit of the first $1/N^2$ correction. Indeed, the reduced
critical coupling seems to saturate towards higher N and one can
estimate that $N=8$ result is within $\approx$ 15\% of the
$N=\infty$ one (a horizontal line).
\begin{figure}[htb]
\vspace{9pt}
\epsfxsize=17pc \epsfbox{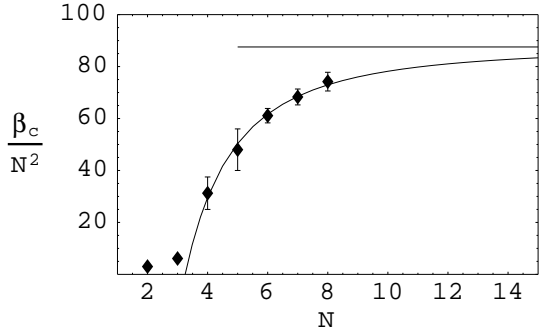} \caption{Dependence of the
normalized transition coupling $\beta_c$ on the number of colours
$N$ at fixed time extension $N_t=4$.} \label{fig:largenenough}
\end{figure}
\section{NONCOMPACT FORMULATION}
We have also studied the new, noncompact formulation of the model
which has better numerical behaviour \cite{APU}. In this approach
the $D-1$ spatial degrees of freedom are noncompact
$X^i(m)=gA^i(m)$ and are defined at the discrete time intervals,
while the temporal one remains compact $U_D(m)=U(m+1,m)$. The
action $S=S_{kin}+S_{pot}$ reads
\begin{equation}
\begin{array}{rcl}
S_{pot}&=&{a\over 2 g^2 }\sum_m  Tr ({\bf X}^i {\bf X}^k)^2, \\
S_{kin} & = &{1\over  a g^2 }\sum_m Tr (\Delta {\bf X}^i) ^2.
\end{array}\label{eq:disc}
\end{equation}
 The  covariant
finite difference along the time direction
\begin{eqnarray}
\lefteqn{\Delta {\bf X}^i(m+1)=}\\
&&{\bf X}^i(m+1)-U(m+1,m){\bf X}^i(m)U(m,m+1),\nonumber
\end{eqnarray}
 takes into account the parallel transport between adjacent lattice sites.
This system has the same local gauge invariance as the compact
version, Eq.(\ref{SG}). With the new action we have extended
previous study to higher dimensions, D=6 and D=10. Preliminary
simulations confirm results found for D=4. In particular the
average size of the system $R^2$ shows a characteristic break in
the temperature dependence at the position consistent with the 't
Hooft scaling. Moreover, the $R^2$ decreases with D in agreement
 with the mean field results \cite{KAB}.
\section{FUTURE PROSPECTS}
Quantitative lattice study of the Yang-Mills quantum mechanics,
and possibly the M-theory, have just begun. Quenched results are
encouraging, but a lot remains to be done. Simulations  work for
all interesting values of the dimension $D$ and are feasible for a
range of N. Recent results give us a rough idea how the large N
limit is approached and where the asymptotics sets it. All
quenched simulations performed up do date indicate existence of
the two regions at finite temperature. This intriguing
correspondence with the predictions of the M-theory should be
further quantified. Of course, the next step is to include the
dynamical fermions. This can be done by a brute force for D=4 and
for the first few N's at D=10.  The one dimensional nature of the
system should help considerably.   For higher N, at D=10, we face
the problem of the complex pfaffian.   An important insight into
the whole subject may be gained by applying the full potential of
the small volume approach \cite{LVB}.


\begin{thebibliography}{9}
\bibitem{BFSS} T. Banks, W. Fishler, S. Shenker and L. Susskind,
 \Journal{\PRD}{55}{6189}{1997}.
\bibitem{CH} M. Claudson and M. B. Halpern, \Journal{\NPB}{250}{689}{1985}.
\bibitem{UPP} U. H. Danielsson, G. Ferretti and B. Sundborg,
\Journal{\IJMPA}{11}{5463}{1996}.
\bibitem{HS} M. B. Halpern and C. Schwartz, \Journal{\IJMPA}{13}{4367}{1998}.
 \bibitem{JW} R. A. Janik and J. Wosiek, 
 \Journal{\APPB}{32}{2143}{2001}; 
 hep-lat/0011031.
\bibitem{STRO} A. Strominger and C. Vafa, \Journal{\PLB}{379}{99}{1996}.
\bibitem{MAR}  E. J. Martinec, hep-th/9909049. 
\bibitem{APU} P. Bialas and J. Wosiek, hep-lat/0109031. 
\bibitem{KAB} D. Kabat, G. Lifszytz and D. A. Lowe, \Journal{\PRL}{86}{1426}{2001}.
\bibitem{LVB} M. L\"{u}scher, \Journal{\NPB}{219}{233}{1983};
 P. van Baal, 
 \Journal{\APPB}{20}{295}{1989};
P. van Baal, hep-ph/0008206.
\end{thebibliography}
\end{document}